\documentclass{article}

\usepackage[preprint,nonatbib]{neurips_2021}

\usepackage[utf8]{inputenc} 
\usepackage[T1]{fontenc}    
\usepackage{hyperref}       
\usepackage{cite}
\usepackage{url}            
\usepackage{booktabs}       
\usepackage{amsfonts}       
\usepackage{nicefrac}       
\usepackage{microtype}      
\usepackage{xcolor}         
\usepackage{bm}

\usepackage{amsthm}
\usepackage{adjustbox}
\usepackage{diagbox}
\usepackage{mathtools}

\usepackage{graphicx} 
\usepackage{caption}
\usepackage{subcaption}
\usepackage{float}
\usepackage{wrapfig}

\newtheorem{Lem}{Lemma}

\title{Principled Hyperedge  Prediction with Structural Spectral Features and Neural Networks}

\author{%
\large Changlin Wan\textsuperscript{\rm 1}, Muhan Zhang\textsuperscript{\rm 2}, Wei Hao\textsuperscript{\rm 1}, Sha Cao\textsuperscript{\rm 3}, Pan Li\textsuperscript{\rm 1} Chi Zhang\textsuperscript{\rm 3}\\
\textsuperscript{1} Purdue University, \textsuperscript{2} Peking University, \textsuperscript{3} Indiana University\\
}

\begin{document}

\maketitle

\begin{abstract}
Hypergraph offers a framework to depict the multilateral relationships in real-world complex data. Predicting higher-order relationships, i.e hyperedge, becomes a fundamental problem for the full understanding of complicated interactions. The development of graph neural network (GNN) has greatly advanced the analysis of ordinary graphs with pair-wise relations. However, these methods could not be easily extended to the case of hypergraph. In this paper, we generalize the challenges of GNN in representing higher-order data in principle, which are edge- and node-level ambiguities. To overcome the challenges, we present \textbf{SNALS} that utilizes bipartite graph neural network with structural features to collectively tackle the two ambiguity issues. SNALS captures the joint interactions of a hyperedge by its local environment, which is retrieved by collecting the spectrum information of their connections. As a result, SNALS achieves nearly 30\% performance increase compared with most recent GNN-based models. In addition, we applied SNALS to predict genetic higher-order interactions on 3D genome organization data. SNALS showed consistently high prediction accuracy across different chromosomes, and generated novel findings on 4-way gene interaction, which is further validated by existing literature. 
\end{abstract}

\section{Introduction}
\begin{figure}
    \centering
    \includegraphics[width=0.9\textwidth]{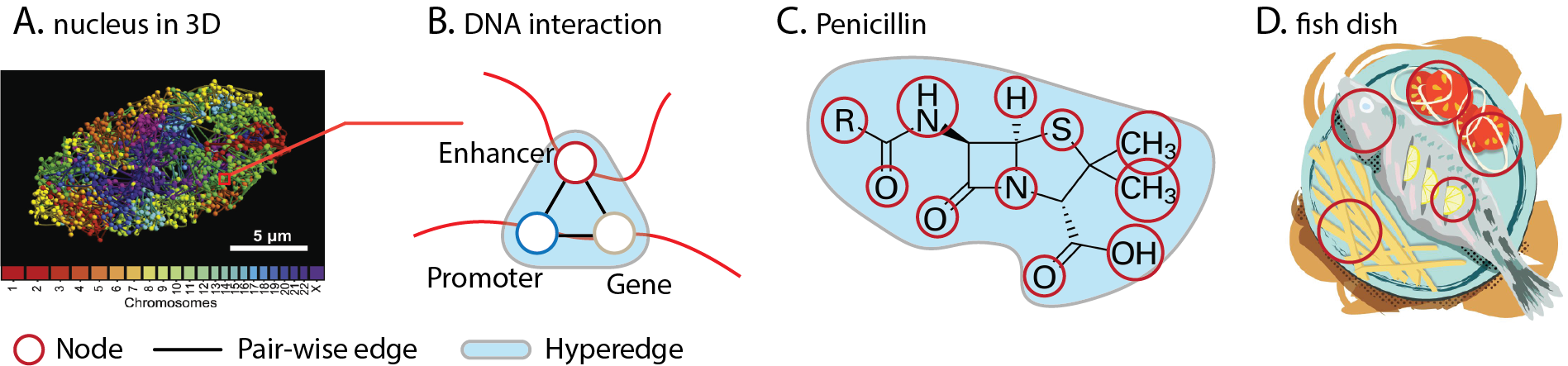}
    \caption{Hyperedge reflects higher-order interaction in many real world data. A. Schematic of cell nucleus in 3D \cite{su2020genome}. B. Illustration of enhancer-promoter-gene for regulated gene expression. C. Molecule diagram of penicillin. D. A flavourful fish dish with multiple ingredients.} 
    \label{fig:DNA}
\end{figure}

Machine learning has been broadly utilized to process graph-structured data in various domains such as e-commence, drug design, social network analysis, and recommendation system \cite{zhou2006learning,gilmer2017neural,stokes2020deep,liben2007link,fan2019graph}. While many methods are devoted to studying ordinary \textit{graphs} that represent pair-wise relations, it has been recognized that many relationships are characterized by more than two participating partners and are thus not bilateral
\cite{feng2019hypergraph,bai2021hypergraph,srinivasan2021learning}. Taking the genetic interaction as an example (figure \ref{fig:DNA}A,B). An accurate characterization of gene expression involves the joint interaction among gene, promoter and enhancer, and capturing only their pairwise interaction (enhancer-promoter, promoter-gene or gene-enhancer) will not fully recapitulate the gene regulatory relationship (figure \ref{fig:DNA}B) \cite{cramer2019organization,sutherland2009transcription,yu2017three}. The same issue also exists in the analysis of multi-component drug design (figure \ref{fig:DNA}C), multi-ingredient recipes (figure \ref{fig:DNA}D), where multilateral relationships are not compatible with ordinary graph edges \cite{jimenez2020drug,yu2016dynamic,zhang2018beyond}. To overcome such conceptual limitations, \textit{hypergraph} has been developed to model the higher-order interaction data \cite{berge1984hypergraphs}. In a hypergraph, any higher-order connection can be represented by a hyperedge that could join an any number of entities (blue shadow figure \ref{fig:DNA}B,C,D). Hence, predicting the higher-order relation is transformed into a hyperedge prediction problem in a hypergraph.

Earlier works on hyperedge prediction rely on structural heuristics~\cite{benson2018simplicial}, such as geometric mean, common neighbor, Adar index~\cite{liben2007link}, as well as higher-order PageRank \cite{nassar2019pairwise}. However, these methods showed unsatisfactory performance due to the limited expressive power of structural heuristics. Recently, graph neural network (GNN) has been introduced as a powerful method for hyperedge prediction, and showed much improved performance \cite{feng2019hypergraph,srinivasan2021learning,yadati2020nhp,bai2021hypergraph,zhang2019hyper}. Essentially, all these methods could be considered as aggregation functions that integrate the information of individual nodes for the representation of hyperedges. However, aggregation-based methods could suffer from severe ambiguities issue in the case of hypergraph. For instance, two hypergraphs could have nodes with identical pairwise connections but different hyperedges that link the nodes (figure \ref{fig:ambiguity}A,B,C). Methods based on the pair-wise node relationship like common neighbor, geometric mean will fail to distinguish such differences. Another example is to consider two different hyperedges whose connected nodes are highly similar (\(\{v_1,v_2,v_3\}\) and \(\{v_2,v_3,v_6\}\) in figure \ref{fig:ambiguity}D.E). Methods like graph neural networks that rely on aggregating the node embedding information for the representation of hyperedges would wrongly predict these two hyperedges to be the same.

In this paper, we generalize the above ambiguity issue as two sub-problems, \textbf{edge-level ambiguity} and \textbf{node-level ambiguity}, which underlie the core differences between ordinary \textit{graphs} and \textit{hypergraphs}. Such size flexibility enables different arrangement of hyperedges. As a result, hypergraph can no longer be bijectively mapped with pair-wise \textit{graph} (edge-level ambiguity). Also unlike two-node edge in \textit{graph}, whose characteristics could largely be explained by aggregating the node information. Whereas in the case of hypergaph, the aggregation scheme could not represent the hyperedge in full as it omits the information of the node dependency within the hyperedge  (node-level ambiguity).

To address these issue for a finer presentation/prediction of hyperedge, we propose \textbf{SNALS} that utilize bipartite graph neural network and node structure feature to compensate such information loss. Compared with most recent models, SNALS achieved nearly 30\% performance increase on the prediction task. We also applied SNALS on higher order genome interaction data, where SNALS showed consistent stability across different chromosomes. Moreover, SNALS gives plausible DNA interaction prediction as the top predicted result is validated by existing literature. 

We summarize our contribution in this work as: 1) We formally describe the current challenges in hyperedge representation learning as edge- and node-level ambiguity, which underlie the core differences between hypergraph with ordinary \textit{graph}. 2) We introduce general frameworks to tackle the ambiguity issues, i.e., bipartite graph neural network for edge-level ambiguity and structure features for node-level ambiguity. 3) Our overall model SNALS achieved around 30\% performance improvement compared with recent state-of-the-art models for hyperedge prediction tasks.


\section{Preliminaries}
\subsection{Notations and mathematical backgrounds}
We first introduce the notations and definitions of hypergraph and relevant mathematical terms used in this work. Denote a set as an uppercase character (e.g. \(X\)), elements in a set as lowercase characters (\(x\)), a vector as a bold lower case character (\(\bm{x}\)), and a matrix as a bold uppercase character (\(\bm{X}\)), respectively. The dimension and indices of entries of a matrix are represented by its upper-script (e.g. \(\bm{X}^{n\times m}\)) and lower-script (e.g. \textit{i}-th row: \(\bm{X}_{i:}\), \textit{j}-th column: \(\bm{X}_{:j}\), and the entry of the \textit{i}-th row and \textit{j}-th column: \(\bm{X}_{ij}\)), respectively. Let \(\mathcal{H}=(V,E)\) represent a hypergraph, where \(V\) is the vertex set \(V=\{v_1,..,v_n\}\) and \(E\) is the edge set \(E=\{S_1,...,S_m\}\), \(E\subseteq P(V)\), and \(P(V)\) represents the powerset of \(V\). The cardinality of a hyperedge \(S\) is defined by the number of nodes in \(S\), which is denoted by \(|S|\). The incidence matrix of a hypergraph is defined as \(\bm{H}\in \{0,1\}^{|V|\times |E|}\), in which \(\bm{H}_{ij}=1\) indicates \(v_i\in e_j\), and otherwise \(\bm{H}_{ij}=0\). Denote a hyperedge representation learning function as \(f:S,\bm{H}\rightarrow \mathbb{R}^k\), \(S\subset V\). The hyperedge prediction problem can be generally formulated as training \(f\) as well as a neural network \(p\) that takes \(f(S,\bm{H})\) as input, by which \(p(f(S,\bm{H}))\) predicts if \(S\) forms a hyperedge. Noted, node representation can be considered as a special case of hyperedge representation, i.e., \(f:S,\bm{H}\rightarrow \mathbb{R}^{k}\), \(S\subset V, |S|=1\).

In this work, we only consider undirected and non-attributed hypergraph, so that the representation learning function $f$ only captures the topological characteristics of the hypergraph. Nevertheless, the method described in this study can be easily extended to the representation learning of hypergraphs with directions and node-/edge-attributes.

\textbf{Definition 2.1 Permutation invariance.} \textit{A permutation operation of a hypergraph is defined by a bijective mapping of its nodes: \(V\rightarrow V\). Denote a permutation as \(\pi\) and the complete set of all \(n!\) such permutations as \(\Pi_n\), where \(n\) is the number of nodes. Denote the permutation operation on any hyperedge \(S\subset V\) as \(\pi(S)=\{\pi(i)|i\in S\}\) and the incidence matrix of the hypergraph after permutation as \(\pi(\bm{H}), \pi(\bm{H})_{ij}=\bm{H}_{\pi(i)j}\).
Similarly, the permutation operation on any hyperedge representation learning function can be defined as \(\pi(f(S,\bm{H}))=f(\pi(S),\pi(\bm{H}))\). A hyperedge representation learning function \(f(S,\bm{H})\) is called permutation invariant if \(\forall S\subset V\), and \(\forall \pi\in \Pi_n\) that only permutes nodes in \(S\), \(f(S,\bm{H})\)=\(f(\pi(S),\pi(\bm{H}))\).}


A good hyperedge representation learning function should be permutation invariant, because otherwise every permutation of nodes might result in a different representation of the same hyperedge.

\begin{figure}
    \centering
    \includegraphics[width=0.8\textwidth]{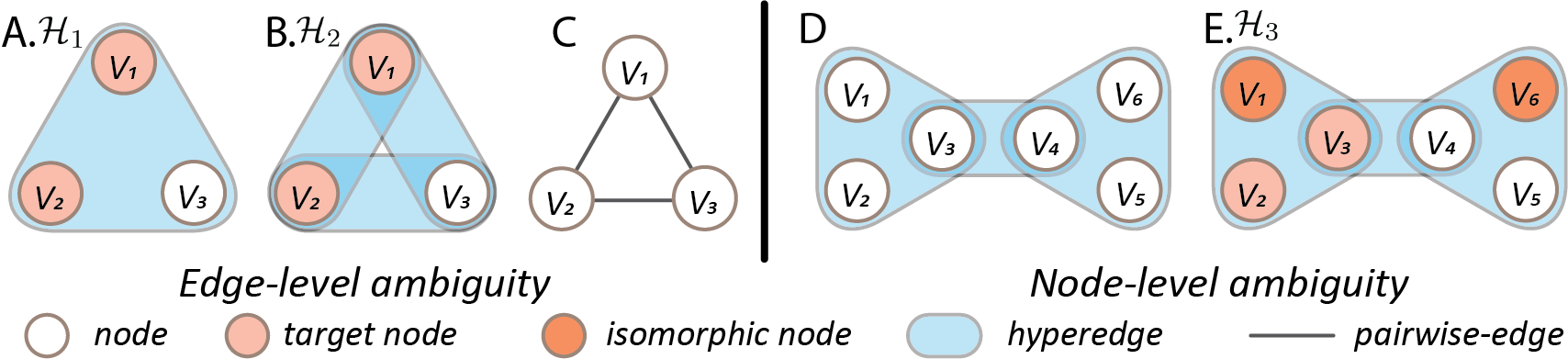}
    \caption{A,B,C. edge-level ambiguity of different hypergraphs that have same pair-wise node connections. D,E. node-level ambiguity of identical nodes \(v_1,v_6\) in different nodes set \(\{v_1,v_2,v_3\},\{v_2,v_3,v_6\}\).}
    \label{fig:ambiguity}
\end{figure}

\textbf{Definition 2.2 Hypergraph isomorphism.}\textit{ Two hypergraphs \(\mathcal{H}=(V,E)\) and \(\mathcal{H}'=(V',E')\) are isomorphic if \(\exists\) a bijective mapping \(\pi:V \rightarrow V', s.t.\ \pi(V)=V'\) and \(\pi(E)=\{\pi(S)|S\subset E\}=E'\), where \(\pi(S)=\{\pi(v)|v\in S\}\). Such a bijective mapping is called an isomorphic mapping. Specifically, a permutation operation \(\pi : V\rightarrow V\) is isomorphic if \(\bm{H}=\pi(\bm{H})\). As isomorphic permutations are exchangeable, we can define the set of all isomorphic permutations of \((V,E)\) as \(\Pi_I\). For any node \(v\), its isomorphic node set is defined by \(I(v)=\{v'|\exists\) \(\pi\in\Pi_I\) s.t. \(\pi(v)=v'\}\), and isomorphic edge set of any edge \(S\) is defined by \(\Pi_I(S)=\{S'|\exists \pi\in\Pi_I\ s.t.\ \pi(S')=S\}\). It is noteworthy that \(\Pi_I\) generates a segmentation of \(P(V)\), denoted as \(\Pi_I(\mathcal{H})\), which can be represented as \(\Pi_I(\mathcal{H})=\{\Pi_I(S_{(i)}) | S_{(i)}\in P(V); \cup \Pi_I(S_{(i)})=P(V); \Pi_I(S_{(i)}) \cap \Pi_I(S_{(j)})= \emptyset, \forall i,j\}\).}

\textbf{Definition 2.3 Isomorphic invariance.} \textit{A hyperedge representation learning function \(f\) is called isomorphic-invariant if for \(\forall S\subset V\) and \(\forall \pi \in \Pi_I\), \(f(S,\bm{H})=f(\pi(S),\pi(\bm{H}))\).}

Isomorphic invariant of \(p\circ f\) is a necessary but insufficient condition of a valid hyperedge predictor, as it ensures a same prediction is made to isomorphic hyperedges. Noted, isomorphic invariant \(f\) is a sufficient but unnecessary condition of the isomorphic invariant property of \(p\circ f\).

\textbf{Definition 2.4 F-invariance.} \textit{Two hypergraphs \(\mathcal{H}=(V,E)\) and \(\mathcal{H}'=(V',E')\) are F-invariant with respect to (w.r.t.) a hyperedge representation learning function \(f\) if \(\exists\) a bijective mapping \(\pi:V \rightarrow V', s.t.\ \forall S\subset V,  f(S,\bm{H})=f(\pi(S),\pi(\bm{H}))\). Specifically, a permutation \(\pi : V\rightarrow V\) is F-invariant w.r.t. \(f\) if \(\forall S\subset V, f(S,\bm{H})=f(\pi(S'),\pi(\bm{H'}))\). We denote the set of all F-invariant permutation as \(\Pi_f\). For a hyperedge \(S\subset V\), we further define the F-invariant edge set of \(S\) as \(\Pi_f(S)=\{S'|\exists \pi\in\Pi_f\ s.t.\ f(\pi(S))=f(S')\}\). Similarly to isomorphic permutations, \(\Pi_f\) also generates a segmentation of \(P(V)\), which is defined by \(\Pi_f(\mathcal{H})=\{\Pi_f(S_{(i)}) | S_{(i)}\in P(V); \cup \Pi_f(S_{(i)})=P(V); \Pi_f(S_{(i)}) \cap \Pi_f(S_{(j)})= \emptyset, \forall i,j\}\).
}
\begin{Lem} Any isomorphic invariant function \(f(S,\bm{H})\) is permutation invariant, and \(\forall S\subset V\), \(\Pi_I(S)\subseteq \Pi_f(S)\).
\end{Lem}

\textit{Proof.} If a hypergraph does not have any non-trivial isomorphic permutation, \(\forall S\subset V\), \(\Pi_I(S)\) only has one element, \(\Pi_I(S)\subseteq \Pi_f(S)\). For a hypergraph having at least one non-trivial isomorphic permutation, and an isomorphic invariant function \(f\), \(f(\Pi_I(S),\Pi_I(\bm{H}))=f(S,\bm{H})\), i.e., S and all of its isomorphic permutations \(\Pi_I(S)\) share the same output of \(f\). Hence \(f\) is permutation invariant and \(\Pi_I(S)\subseteq \Pi_f(S)\).


\begin{Lem}
For a hyperedge \(S\subset V\), if \(\exists\) a permutation \(\pi\ s.t. f(S,\bm{H})=f(\pi(S),\pi(\bm{H}))\) and \(f\) is isomorphic invariant, then \(\pi(S)\in \Pi_f(S)\).
\end{Lem}

\textit{Proof.} Considering the bijective mapping \(\pi_0: \pi_0(S)=\pi(S), \pi_0(\pi(S))=\pi^{-1}(\pi(S))=S\), and \(\pi_0(v)=v, v\in V\setminus (S\cup\pi(S))\), \(\pi_0\) is a permutation. Since \(f(S,\bm{H})=f(\pi(S),\pi(\bm{H}))\) and \(f\) is permutation invariant, \(f(A)=f(\pi_0(A)),\ \forall A\subset (S\cup\pi(S))\). And \(\pi_0\) is an identical mapping for \(v\in V\setminus (S\cup\pi(S))\). Hence, \(\pi_0\) is an F-invariant permutation w.r.t. \(f\), i.e., \(\pi(S)\in \Pi_f(S)\).


Lemma 1 suggests that the segmentation \(\Pi_I(\mathcal{H})\) is always finer than \(\Pi_f(\mathcal{H})\). Because \(p(f(S,\bm{H}))\) has the same output for \(S\subset \Pi_{p\circ f}(S)\), the F-invariant edge set of representation learning function \(\Pi_{p\circ f}(S)\) is expected to be close to the isomorphic set of hyperedges, i.e., \(\Pi_{p\circ f}(S)\approx \Pi_I(S)\). Specifically, \(\Pi_{p\circ f}(S)\setminus \Pi_I(S)\) could reflect the ratio of false-discoveries in distinguishing \(\Pi_I(S)\). For two permutation invariant functions \(f_1\) and \(f_2\), if \(\forall S\subset V\), \(\Pi_{f_1}(\mathcal{S})\subseteq \Pi_{f_2}(\mathcal{S})\), we call \(f_1\) is more informative than \(f_2\) in representing \(\mathcal{H}\). Lemma 2 characterizes a general condition of the hyperedges in a same F-invariant edge set.

\subsection{Related works}
Recent development of graph neural network (GNN) on representation learning tasks achieved unprecedented performance \cite{zhang2018link,kipf2016semi,hamilton2017inductive}. The representation learning function of GNN takes a general form as \[\bm{X}^{l+1}=\sigma(\bm{D}^{-1/2}\bm{A}\bm{D}^{-1/2}\bm{X}^{l}\bm{W}^l)\], where \(\bm{X} \in \mathbb{R}^{|V|\times k}\) represents the learned node embedding, \(\sigma\) is a non-linear activation function, \(\bm{A} \in \{0,1\}^{|V|\times |V|}\) is the adjacency matrix of the input graph. In hypergraph embedding, the adjacency matrix is defined by a clique expansion of the incidence matrix, \(\bm{A}=sign(\bm{H}\bm{H}^T)\in\{0,1\}^{|V|\times |V|}\), in which \(\bm{A}_{ij}=1\) if node \(v_i\) and \(v_j\) belong to at least one hyperedge, and otherwise, \(\bm{A}_{ij}=0\) \cite{yang2020hypergraph}. \(\bm{D}_{ii}=\sum_{j}\bm{A}_{ij}\) is the degree matrix and \(\bm{W}^l\in \mathbb{R}^{k\times k}\) is the layer-specific weight matrix for the \(l\)th layer. 

Noting the adjacency matrix loses substantial topological characteristics of a hypergraph, Hypergraph GNN (HGNN) was then developed by utilizing the incidence matrix \(\bm{H}\) to replace \(\bm{A}\) in the representation learning function \cite{feng2019hypergraph,bai2021hypergraph}, i.e. \[\bm{X}^{l+1}=\sigma(\bm{D}_v^{-1/2}\bm{H}\bm{W}_E^l\bm{D}_e^{-1}\bm{H}^TD_v^{-1/2}\bm{X}^{l}\bm{W}_V^{l})\]
, where \(\bm{D}_{V_{ii}}=\sum_{j}\bm{H}_{ij},\bm{D}_{E_{jj}}=\sum_{i}\bm{H}_{ij}\) are the degree matrix for nodes and hyperedge, respectively. \(\bm{W}_E\in \mathbb{R}^{|V|\times|V|}\) is a diagonal matrix with edge weights as its diagonal elements. This approach along with its variants could be regarded as a weighted clique expansion, i.e, \(\bm{H}\bm{W}_E\bm{D}_e^{-1}\bm{H}^T\approx \bm{W'}\bm{A}\), where \(\bm{W}'\) is a weighting matrix. This formulation still ignores a large amount of topological information of a hypergraph \cite{agarwal2006higher}. 

To further improve the solution, recent works treats each hyperedge as a specific "node" object \cite{dong2020hnhn,arya2020hypersage}, which iteratively learns edge and node representation functions as
\[\bm{X}_E^{l+1}=\sigma(\bm{H}^T\bm{X}_V^l\bm{W}_E^l) \quad and \quad \bm{X}_V^{l+1}=\sigma(\bm{H}\bm{X}_E^l\bm{W}_V^l)\]
, where \(\bm{X}_E\in \mathbb{R}^{|V|\times k}\) is the learned edge embedding. \(\bm{W}_E^l, \bm{W}_V^l\in \mathbb{R}^{k\times k}\) are the layer-specific weight matrix for node\(\rightarrow\)edge/edge\(\rightarrow\)node information transformation. 

Moest recently, Srinivasan et al developed a hyperedge family (FamilySet) based representation learning function. In addition to above methods, FamilySet updates node-/edge-wise embeddings with their nearby nodes/edges, whose representation function follows \cite{srinivasan2021learning}
\[\bm{X}_E^{l+1}=\sigma(concat(\bm{A}_E\bm{X}_E^l,\bm{H}^T\bm{X}_V^l)\bm{W}_E^l)\quad and \quad \bm{X}_V^{l+1}=\sigma(concat(\bm{A}_V\bm{X}_V^l,\bm{H}\bm{X}_E^l)\bm{W}_V^l)\]
, where \(\bm{A}_V\) is the clique expansion that records nearby nodes information and \(concat\) represents concatenation. \(\bm{A}_E\) is the line graph for connected hyperedges \cite{tyshkevich1996line}, where \(\bm{A}_{E_{ij}}=1\) if \(\exists v\in V\) s.t \(v\in S_i, v\in S_j\), otherwise \(\bm{A}_{E_{ij}}=0\). Utilization of the line graph enhanced the communication between hyperedges with its local environment thus enables the tasks like hyperedge expansion.

\subsection{Limitations of existing approaches}
\textbf{Edge-level ambiguity} is defined by a false assessment of identical node embedding to non-isomorphic nodes. For example, the representation learning functions of GNN and HGNN could be simplified as \(f(S,\bm{A})\). Adjacency matrix  over-simplifies the topological characteristics of a hypergraph, which can cause an edge-level ambiguity as showcased in Figure \ref{fig:ambiguity}A,B,C. Clearly, any two nodes from two sub-hypergraphs \(\mathcal{H}_1=\{V_1,E_1\}\) and \(\mathcal{H}_2=\{V_2,E_2\}\) are not isomorphic. However, due to \(\mathcal{H}_1\) and \(\mathcal{H}_2\) have the same clique expansion \(\bm{A}_{\mathcal{H}_1}=\bm{A}_{\mathcal{H}_2}\), GNN and HGNN assign the same node embedding for any nodes from them. Hence, for \(\forall S_1\subset V_1,\ S_2\subset V_2\), and \(S_1\) and \(S_2\) of the same cardinality, GNN and HGNN have \(p(f(S_1))=p(f(S_2))\), i.e. \(S_2\in \Pi_{p\circ f}(S_1)\) and \(S_2\not\in \Pi_I(S_1)\).

\textbf{Node-level ambiguity} is defined by a false assessment non-isomorphic edges with an identical embedding. Although edge embedding \(\bm{X}_E\) was introduced in recent works \cite{arya2020hypersage,srinivasan2021learning,dong2020hnhn}, it was only computed for existing edges in \(\mathcal{H}\) and served as an intermediate step to update node embedding \(\bm{X}_V\). Actually, all the aforementioned methods adopt an aggregation (e.g. sum-pooling) of node embedding when predict unknown hyperedges, i.e., 
\[f(S,\bm{H})=AGG(f(v_1,\bm{H}),f(v_2,\bm{H}),...,f(v_m,\bm{H})|v_1,...,v_m\in S), \forall S\subset V\]

\begin{Lem} For any isomorphic invariant edge representation learning function follows  \(f(S,\bm{H})=AGG(f(v_1,\bm{H}),f(v_2,\bm{H}),...,f(v_m,\bm{H})), v_1,...,v_m\in S\), and \(\forall S'=\{v_1'\in\Pi_f(v_1), v_2'\in\Pi_f(v_2), ... , v_m'\in\Pi_f(v_m)\}\), then \(S'\in\Pi_f(S)\).
\end{Lem}

\textit{Proof.} As \(f\) is isomorphic invariation, \(f(S,\bm{H})=f(S',\bm{H})\), and by Lemma 2, \(S'\in\Pi_f(S)\).


A simple aggregation of node embedding ensures a high computational feasibility and a easy handling of the edges of different cardinality. However, \textbf{Lemma 3} suggests that isomorphic invariant \(f\) ignores the topological dependency of nodes within \(S\) when it adopts the aggregation based formulation, i.e. \(f(S,\bm{H})\perp \bm{H}|\{f(v_1,\bm{H}),f(v_2,\bm{H}),...,f(v_m,\bm{H})\}\). Hence, all aforementioned methods suffer an over simplified edge embedding. Figure \ref{fig:ambiguity}D,E illustrated one example of node-level ambiguity caused by such over-simplification. In the hypergraph, \(\{v_1, v_2, v_5, v_6\}\) are isomorphic and \(\{v_3, v_4\}\) are isomorphic. If \(f\) satisfies Lemma 3, \(f(v_1)=f(v_5)\), then  \(p(f(v_1,v_2,v_3))=p(AGG(f(v_1),f(v_2),f(v_3)))=p(f(v_1,v_2,v_56))\). However, the node sets \(S_1=\{v_1,v_2,v_3\}\) and \(S_2=\{v_2,v_3,v_5\}\) clearly have different topological structure, i.e. \(S_2\in \Pi_{p\circ f}(S_1)\) and \(S_2\not\in \Pi_I(S_1)\).

\begin{figure}
    \centering
    \includegraphics[width=0.7\textwidth]{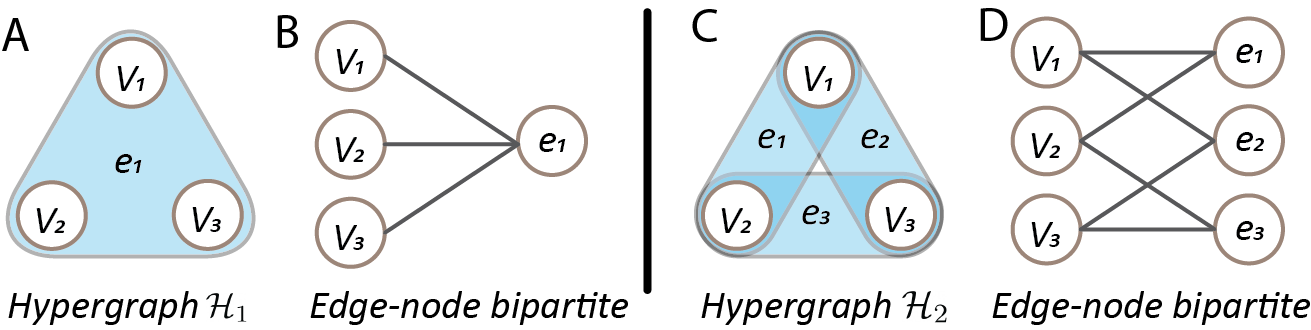}
    \caption{\textbf{Edge-level ambiguity} is caused by different hyperedges with same clique expansion.}
    \label{fig:edge}
\end{figure}

\section{Methodology}
\vspace{-2mm}
In order to avoid the edge-level ambiguity, a desired representation function should embed each single node by encoding sufficient topological characteristics of the hyperedges containing that node. \textbf{Lemma 3} suggests that the node-level ambiguity is inevitable if \(f\) is isomorphic invariant. Here we developed SNALS to effectively alleviate these two types of ambiguities by (1) adopting a bipartite message passing neural network, (2) encoding local topological characteristics in hyperedge representation learning function, and (3) utilizing the spectrum characteristics of each node set in hyperedge prediction. The mathematical considerations of SNALS include (i) utilizing local topological characteristics and (ii) introducing spectrum characteristics in hyperedge representation, which keeps the isomorphic invarint property of \(f\) but changes the aggregation form. These two considerations leverage computational feasibility and integration of node set specific topological characteristics in hyperedge prediction.

\subsection{Bipartite message passing neural network.}
Considering each hyperedge as an individual object, the hypergraph \(\mathcal{H}=(V,E)\) could be manifested as a bipartite graph, where one partite represents nodes \(V\) and the other represents the hyperedges \(E\) (figure \ref{fig:edge}). The edge-node bipartite graph is equivalent to the incidence matrix \(\bm{H}\), which conceive more information than the clique expansion \(\bm{A}=sign(\bm{H}\bm{H}^T)\). 
Bipartite message passing neural networks have been utilized in previous studies \cite{arya2020hypersage,dong2020hnhn}. In SNALS, we utilize a modified Bipartite message passing neural networks by introducing a one-side normalization term \(\bm{D}_E^{-1}\) when updating \(\bm{X}_E\), 
\[\bm{X}_E^{l+1}=\sigma(\bm{H}^T\bm{X}_V^l\bm{D}_E^{-1}\bm{W}_E) \quad and \quad \bm{X}_V^{l+1}=\sigma(\bm{H}\bm{X}^l_E\bm{W}_V)\]
Noted, the nonlinear activation in updating \(\bm{X}_E\) and \(\bm{X}_V\) enables a flexible and optimized information retrieval from \(\bm{H}\), which is more informative than a clique expansion based representation, i.e. 
\[\bm{H}\sigma(\bm{H}^T\bm{X}_E\bm{W}_E)\bm{W}_V\neq \bm{W}'\bm{A}\bm{X}_V\bm{W}_V\] The representation function of bipartite message passing neural network distinguishes edges of different cardinality, which is more practical in modeling real word hypergraphs. In addition, the one-sided normalization approach would balance the trade off between degree bias and representation power. Our experiments suggested that the one-sided normalization has a better performance than  normalizing both \(\bm{X}_V\) and \(\bm{X}_E\) or none normalization (Appendix). Specifically, the bipartite message passing framework can effectively handle edge-level ambiguity as it consider the full topological characteristics, \(\bm{H}\), when representing each node.

\begin{figure}
    \centering
    \includegraphics[width=0.9\textwidth]{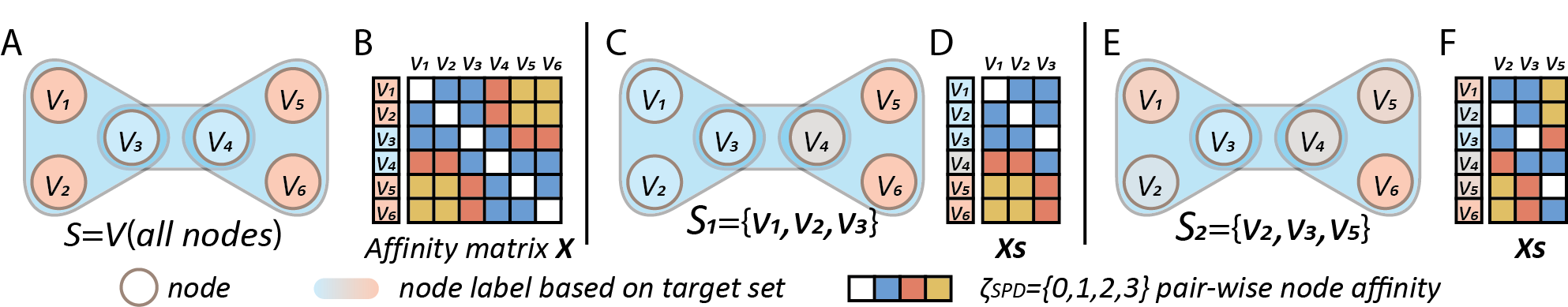}
    \caption{\textbf{A.B.} In terms of whole hypergraph, \(v_1,v_2,v_5,v_6\) and \(v_3,v_4\) are isomorphic. The isomorphism is also reflected by affinity matrix \(\bm{X}\), i.e., for two isomorphic nodes \(v_i,\,v_j\), \(\exists \pi, s.t\, \bm{X}_{i:}=\pi(\bm{X}_{j:})\). \textbf{C.D.E.F.} The isomorphic property of nodes changed by focusing on the relationship with target nodes set \(S\). The \(S\)-specific affinity matrix \(\bm{X}_{S_{(q)}}\) is thus regarded as the structure feature of nodes for the topological representation of hyperedge \(S\).}
    \label{fig:structure}
\end{figure}

\subsection{Hyperedge representation by using structural  features}
\textbf{Definition 3.1 Hyperedge local representation.} \textit{Given a target node set \(S\), its \(q\)-hop neighbor nodes set \(V_{(q)}\), edges set \(E_{(q)}\), incidence matrix \(\bm{H}_{(q)}\) and affinity matrix \(\bm{X}_{{S}_{(q)}}\) are defined as follows: \(V_{(q)}=\{v_j|\zeta_{SPD}(v_i,v_j)\leq q,\forall v_j \in V, \forall v_i \in S\}\). \(E_{(q)}=\{e_i|e_i\subseteq V_{(q)} , \forall e_i\in E\}\). \(\bm{H}_{(q)}\in \{0,1\}^{|V_{(q)}|\times |E_{(q)}|}\), where \(\bm{H}_{{(q)}_{ij}}=1\) if \(V_{q_i}\in E_{q_j}\), otherwise \(\bm{H}_{q_{ij}}=0\). \(\bm{X}_{{S}_{(q)}}\in \mathbb{R}^{|V_{(q)}|\times |S|}\), where \(\bm{X}_{{S}_{{(q)}_{ij}}}=\zeta_{SPD}(v_i,v_j|v_i\in V_{(q)}, v_j\in S)\) represents the shortest path distance between \(v_i\) and \(v_j\).}

The definition of \(q\)-hop neighbor nodes set \(V_{(q)}\), edges set \(E_{(q)}\), and incidence matrix \(\bm{H}_{(q)}\) are intuitive. The affinity matrix \(\bm{X}_{{S}_{(q)}}\) is defined by the distance matrix of the nodes in the \(q\)-hop neighbor of \(S\), where the distance is shortest path distance \(\zeta_{SPD}\). Straightforwardly, for any isomorphic permutation \(\pi\), 
\(\pi(\bm{H}_{(q)})=\bm{H}_{(q)}\) and \(\pi(\bm{X}_{S_{(q)}})=\bm{X}_{{S}_{(q)}}\).  
Hence, both \(\bm{H}_{(q)}\) and \(\bm{X}_{{S}_{(q)}}\) are valid inputs of a isomorphic invariant representation function. Figure \ref{fig:structure} illustrates the affinity matrices of the hop-1 neighbors of the complete hypergraph (Figure \ref{fig:structure}A, B), the node set \(S_1=\{v_1, v_2, v_3\}\) (Figure \ref{fig:structure}C, D) and 
\(S_2=\{v_2, v_3, v_5\}\) (Figure \ref{fig:structure}E, F). Noted, by \textbf{Lemma 3}, any isomorphic invariant \(f(S)=AGG(f(v|v\in S))\), \(f(S_1)=f(S_2)\) as \(f(v_1)=f(v_5)\). On the other side, as \(\bm{X}_{S_{1_{(1)}}}\neq \bm{X}_{S_{2_{(1)}}}\), instead of a simple aggregation of their elements, the structural difference of \(\bm{X}_{S_{1_{(1)}}}\) and \(\bm{X}_{S_{2_{(1)}}}\) can distinguish \(S_1\) and \(S_2\). Based on this idea, we propose a new hyperedge representation learning function by using the structural information of \(\bm{X}_{{S}_{(q)}}\), which is isomorphic invariant but does not follow an aggregation form. 

Encoding the structural information of affinity matrix causes addition computational complexity. To ensure the computational feasibility, instead of the representation with entire graphs, hyperedge local representation only requires \(q\)-hop enclosing subgraph \(V_{(q)}\) around \(S\). Practically, \(q\leq 2\) is sufficient for a good prediction performance. We argue that  \(q\leq 2\) is a practical setting in real-world analysis, because (1) exact isomorphic nodes are rare in real-world hypergraphs, utilizing local information to determine similar nodes can increase the inductive power of the model; (2) For the task of hyperedge prediction, the nodes beyond 2-hop are less deterministic for the existence of a hyperedge; (3) \(q\leq 2\) bounds the size of \(\bm{H}_{(q)}\) and \(\bm{X}_{{S}_{(q)}}\) that dramatically reduce the computational cost and can be directly implemented in the message passing neural network. We present our model rooted in \(f(S,\bm{X}_{S_{(q)}},\bm{H}_{(q)})\) for the hypergraph edge representation/prediction task in the next section.

\begin{figure}
    \centering
    \includegraphics[width=0.85\textwidth]{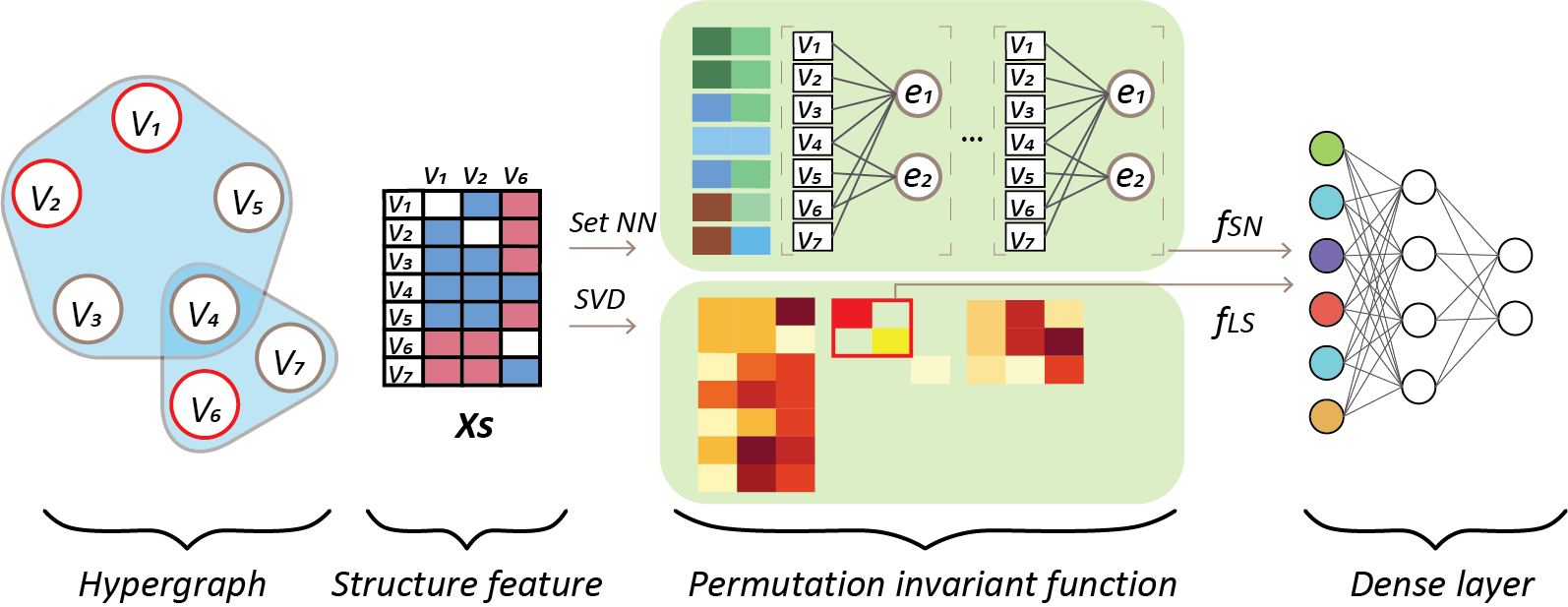}
    \caption{\textbf{The SNALS framework}. To predict the existence of hyperedge \(S\), SNALS tackles edge-/node-level ambiguity by integrating bipartitte graph neural network with structure feature. To captures the joint connections of target nodes with its nearby nodes, SNALS retrieves the local spectrum information of structure feature matrix. Followed by a dense layer, SNALS combines all the information and gives the prediction result.}
    \label{fig:model}
\end{figure}

\subsection{SNALS for hyperedge prediction}

We represent our model \textbf{SNALS} (\textbf{S}tructural representing \textbf{N}eural network \textbf{A}nd \textbf{L}ocal \textbf{S}pectrum) as the integration of two isomorphic invariant functions \(f_{SN}(S,\bm{H}_{(q)},\bm{X}_{S_{(q)}})\) and \(f_{LS}(S,\bm{X}_{S_{(q)}})\) for hyperedges representation learning. 

\textbf{Structural representing neural network}

A series of work has explored the fixed edge representation with affinity matrix \(\bm{X}_{S_{(q)}}\in \mathbb{R}^{|V_{(q)}|\times k}\), which grants new labels for each node in the presentation of edge \(S\) by imposing a permutation invariant function on every row of \(\bm{X}_{S_{(q)}}\) (figure \ref{fig:structure}C,E). For instance, to predict the link in graph, i.e., \(k=2\), SEAL considered a specific type of node labeling by tracking distances of a node to the target two nodes and showed superior performance over existing methods \cite{zhang2018link,zhang2020revisiting}. Li et al. further generalized such a definition to the case with \(S\) of arbitrary sizes but they still work on graphs instead of hypergraphs \cite{li2020distance}. Mathematically, Li et al. characterized the expressive power of the obtained GNNs  which solve the edge-level ambiguity issue previously observed in graphs \cite{srinivasan2019equivalence}. Motivated by these works, we propose structural representing representation learning function \(f_{SN}\), which integrates affinity matrix \(\bm{X}_{S_{(q)}}\) by using a bipartite graph neural network. Unlike \(k\)-size edge representation, the affinity matrix \(\bm{X}_{S_{(q)}}\in \mathbb{R}^{|V_{(q)}|\times |S|}\) has varied dimension depending on the size of hyperedge \(S\). To construct a uniformed input, we first process \(\bm{X}_{S_{(q)}}\) by using a set neural network (setNN) developed in precisely deepsets \cite{zaheer2017deep} (see details in  Appendix). Specifically, by treating each row of \(\bm{X}_{S_{(q)}}\) as an individual set vector, setNN acted as a permutation invariant function to standardize the node-wise feature into a feature matrix of a fixed dimension, i.e., \(\bm{X}_{(q)}^0=f_{setNN}(\bm{X}_{S_{(q)}}), \bm{X}_{(q)}^0\in \mathbb{R}^{|V_{(q)}|\times k}\). This feature matrix is thus served as the input node features \(\bm{X}_{V_{(q)}}^0\) to initiate the message passing in the bipartite neural network:
\[\bm{X}_{V_{(q)}}^{l+1}=\sigma(\bm{H}\bm{X}^l_{E_{(q)}}\bm{W}_V^l),\quad \bm{X}_{E_{(q)}}^{l+1}=\sigma(\bm{H}^T\bm{X}_{V_{(q)}}^{l}\bm{D}_E^{-1}\bm{W}_E^l),\quad \bm{X}_{V_{(q)}}^0=f_{setNN}(\bm{X}_{S_{(q)}})\]
Noted, \(f_{SN}\) also follows the form of aggregation, i.e., \(f_{SN}(S)=AGG(f_{SN}(v|v\in S))\).


\textbf{Spectrum of the local structure feature matrix}

As \(f_{SN}\) still follows the aggregation form, by \textbf{Lemma 3}, it still suffers node-level ambiguity. In sight of this, we propose \(f_{LS}\) that keeps the structure of the affinity of \(\bm{X}_{S_{(q)}}\) in the representation of \(S\). Specifically, \(f_{LS}\) is a function based on the singular values \(\bm{X}_{S_{(q)}}\), i.e., the spectrum of the subgraph \(S_{(q)}\). 
The rationale is that singular values reflects the low rank property of the affinity matrix \(\bm{X}_{S_{(q)}}\), i.e., the topological structure of \(S_{(q)}\). Intuitively, the affinity matrix with a higher low-rankness suggests the nodes in \(S\) are of higher topological similarity. Noted, as the singular value decomposition is invariant to row and column wise shuffles, \(f_{LS}\) based on the singular values of \(\bm{X}_{S_{(q)}}\) is also isomorphic invariant. However, \(f_{LS}\) does not follow the aggregation form. To cope with hyperedge (\(|S|\geq 2\)) of varied sizes, \(f_{LS}\) only takes the two largest singular value into account, the ratio between is sufficient to characterize the level of low-rankness of \(\bm{X}_{S_{(q)}}\), 
\[f_{LS}(S,\bm{X}_{S_{(q)}})=f(\bm{\Sigma}_{11},\bm{\Sigma}_{22}), \quad \bm{X}_{Sq}=\bm{U}\bm{\Sigma}\bm{V}^T\]

Together, we present the \textbf{SNALS} framework (figure \ref{fig:model}) that integrates the \textbf{S}tructure \textbf{N}eural network \(f_{SN}\) \textbf{A}nd \textbf{L}ocal \textbf{S}pectrum information of structural feature matrix \(f_{LS}\). Follows by a dense layer, SNALS gives the prediction for hyperedges by \(p(concat(f_{SN}(S,\bm{H}_{(q)},\bm{X}_{S_{(q)}}),f_{LS}(S,\bm{X}_{S_{(q)}})))\).

Noted, as \(f_{SN}(S,\bm{H}_{(q)},\bm{X}_{S_{(q)}})\) also follows the aggregation form, for any other representation learning function \(f\) satisfies the condition of \textbf{Lemma 3}, the partition generated by the F-invariant edge sets of \(f_{SN}\) and \(f\), i.e., \(\Pi_{f_{SN}}(\mathcal{H})\) and \(\Pi_{f}(\mathcal{H})\) can be easily compared. Specifically, if \( \Pi_{f}(\mathcal{H})\subset \Pi_{f_{SN}}(\mathcal{H})\), we can replace \(f_{SN}\) by \(f\) in SNALS. By \textbf{Lemma 3}, the \(\Pi_{p_{SNALS}}(concat(f(S,\bm{H}),f_{LS}(S,\bm{X}_{S_{(q)}}))(S)\subset \Pi_{p(f(S,\bm{H}))}(S)\), i.e., the SNALS framework can always enable a finer representation to the representation learning function \(f\) satisfies the condition of \textbf{Lemma 3}.


\section{Experiments}

\begin{table}
  \label{sample-table}
  \begin{adjustbox}{max width=\linewidth,center}
  \begin{tabular}{lcccccccc}
    \toprule
   \backslashbox{Methods}{Data} & DAWN & email-Eu & NDC-class & NDC-substance & threads-ask & threads-math & tags-ask & tags-math \\
    \midrule
    HGNN  & 0.624(0.010) &  0.664(0.003) & 0.614(0.005) & 0.421(0.014) & 0.425(0.007) & 0.453(0.007) & 0.545(0.005) & 0.599(0.009) \\
    \(\bm{H}\)RGCN    & 0.634(0.003) & 0.661(0.006) & 0.676(0.049) &  0.525(0.006) & 0.464(0.010) & 0.487(0.006) & 0.545(0.006) & 0.572(0.003)  \\
    FamilySet   & 0.677(0.004) & 0.687(0.002) & 0.768(0.004) & 0.512(0.032) & 0.605(0.002) & 0.586(0.002) & 0.605(0.002) & 0.642(0.006) \\
    SetSEAL &  0.814(0.013) & 0.758(0.011) & 0.822(0.015) & 0.868(0.019) & 0.581(0.015) & 0.483(0.021) & 0.798(0.018) & 0.833(0.015)  \\
    \midrule
    SNALS(\(f_{SN}\)) & \textbf{0.840(0.012)} & 0.780(0.010) & 0.880(0.021) & 0.914(0.007) & 0.623(0.024) & 0.627(0.018) & \textbf{0.823(0.009)} & \textbf{0.869(0.006)} \\
    SNALS(\(f_{SN},f_{LS}\)) & \textbf{0.838(0.010)} & \textbf{0.785(0.011)} & \textbf{0.896(0.020)} & \textbf{0.918(0.006)} & \textbf{0.714(0.019)} & \textbf{0.654(0.027)} & \textbf{0.822(0.012)} & \textbf{0.869(0.006)}\\
    \bottomrule
  \end{tabular}
  \end{adjustbox}
\caption{Model performance comparison on F1 score for different hypergraph data.}
\end{table}

In this section, we briefly introduce the experimental setup and benckmark datasets, and  evaluate the hyperedge prediction accuracy of SNALS with state-of-the-art (SOTA) methods using F-1 score.

\textbf{Baseline methods.} Our first baseline method HGNN is based on clique expansion, and is expected to be affected by both ambiguities. The second baseline method employes relational graph neural network on node-edge bipartite expansion (\(\bm{H}\)RGCN) to offset edge-level ambiguity \cite{srinivasan2021learning}, but not node-level ambiguity. For our third baseline method, we access methods that deal with node-level ambiguity but are affected by edge-level ambiguity. As reflected in figure \ref{fig:structure}, the node level ambiguity is rooted from the disagreement between affinity matrix \(\bm{X}\) and structural feature \(\bm{X_S}\). A series of work try to amend such differences in edge presentation by adding additional affinity labels to each nodes\cite{zhang2018link,zhang2020revisiting,li2020distance,srinivasan2019equivalence}. Among them, SEAL is the SOTA algorithm in utilizing structure feature as node label for link prediction\cite{zhang2018link}. To adopt SEAL in hyperedge prediction, we propose setSEAL as our third baseline that replace the node labeling approach in SEAL with deepsets. SetSEAL is expected to be affected by edge-level ambiguity. The fourth method FamilySet adopts the ideas of both clique expansion and line graph \cite{srinivasan2021learning}, and is expected to avoid edge-level ambiguity and partially node-level ambiguity. For our model SNALS, we also compared the performance with or without the added local spectrum information \(f_2\) to illustrate the necessity to add the spectrum information.

\textbf{Datasets \& Evaluation.} Eight hypergraph datasets\footnote{All data are retrieved from www.cs.cornell.edu/~arb/data/} were utilized in our evaluation, namely question tags of stock exchange forum(tag-ask, tag-math), user question answering in online threads (thread-ask, thread-math), different component of drugs (NDC-classes, NDC-substances), combined drug use by a patient (DAWN) and emails with recipient addresses (email-Eu). For each dataset, we keep only the hyperedges containing at least two nodes. Detailed information of the processed datasets are in appendix. Following \cite{srinivasan2021learning}, for each dataset, we generate 5 times negative hyperedges to the real ones as negative training data.

\textbf{Results.} Using 5-fold cross validation, we report the mean and standard deviation of the F1-scores for all methods in Table 1. setSEAL showed better performance than other baseline methods, indicating the more profound impact of node-level ambiguity over edge-level ambiguity. Similarly for the better performance of FamilySet over \(\bm{H}\)RGCN. By tackling both node- and edge-level ambiguity, SNALS on average achieves 30\% performance increase over FamilySet and 13\% performance increase over setSEAL. Compared with SNALS(\(f_1\)), adding the local spectrum information, namely, SNALS(\(f_1\)), further increases model performance, especially for the threds data. We also empirically showed that the local spectrum information is indispensable in appendix. To maximize the power of SNALS, we extensively tested its hyperparameters and gives recommendations for each component of SNALS in appendix. Taken together, SNALS achieved great representation power for the edge prediction task.
\vspace{-2mm}
\section{Hyperedge prediction on DNA interactions data}
\vspace{-2mm}
In mammalian cells, the 3D genome organization is proven to function in many biological processes, and the higher-order chromatin organizations are frequently linked to long-distance gene regulation that could control development and cell fate commitment \cite{dixon2015chromatin,lee2019simultaneous}. As introduced earlier, genetic interactions are higher-order connections that involve multiple entities, such as gene, enhancer, promoter, et al \cite{cramer2019organization,sutherland2009transcription,yu2017three}. Current methods for analyzing the genome organization data are still limited to pair-wise connections, while efficient tools/methods are lacking for the exploration of higher order interactions 
in 3D genome data \cite{quinodoz2018higher,beagrie2017complex,tavares2020multi}.

\begin{figure}
    \centering
    \includegraphics[width=0.75\linewidth]{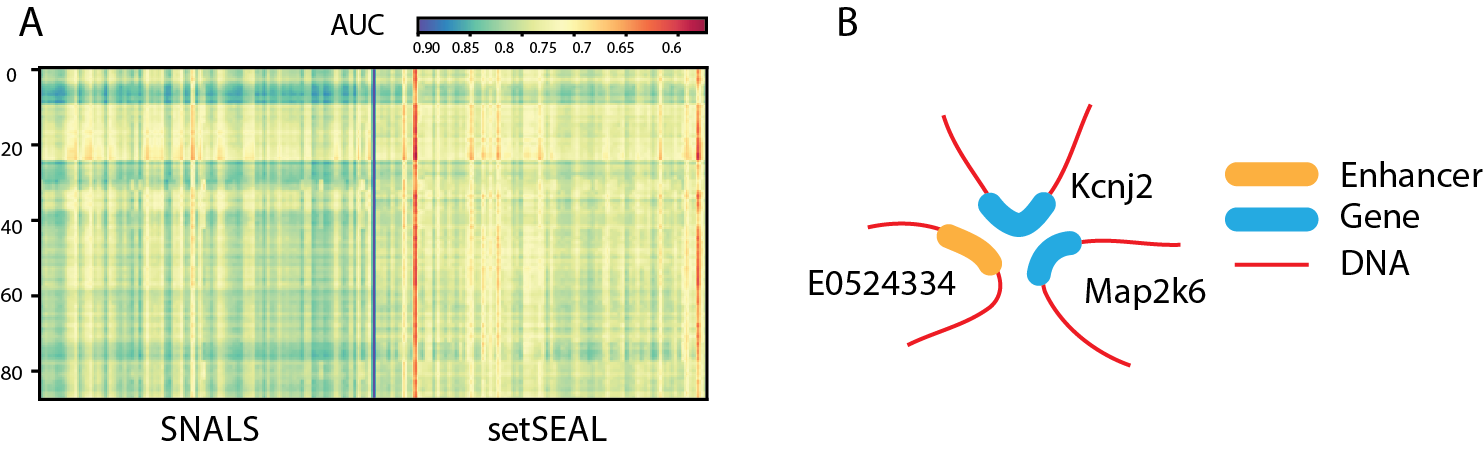}
    \caption{SNALS gives plausible prediction of higher order genetic interaction.}
    \label{fig:sprite}
\end{figure}

As a proof of concept study, here we utilize SNALS to predict the genome higher-order interactions (hyperedge) of mouse embryonic cells. We retrieve the 3D genome connection data from \cite{quinodoz2018higher}. Similarly, we only keep the hyperedges that have at least two nodes, and constructed the negative training data by generating five negative hyperedges for each hyperedge observed. We first test whether our model could achieve consistent performance across different chromosomes. For each of the 17 
autosomals in mouse genome, we randomly selected 5 autosomals to study the interactions, resulting in \(85\times 85\) pair-wise cross validations. We compared SNALS with the strongest baseline method setSEAL and report the Area under ROC curve (AUC) in figure \ref{fig:sprite}A. In general, SNALS outperforms setSEAL across all the test conditions. More importantly, the performance of SNALS is very stable, since it did not show any bias towards specific chromosomes, unlike setSEAL on chromosome 2 and 17.

We then apply SNALS to predict the 4-way genetic interactions in chromosome 11. One hyperedge corresponding to the interactions of the bin elements 5521, 5589, 5602, 5630, is predicted by SNALS that is not captured by original assay. These bins are located within the same topological associated domain (TAD) \cite{dixon2012topological}. Furthermore, we find genes Map2k6, Kcnj2 and enhancer E0524334 are located within 5521, 5589, 5602, respectively (figure \ref{fig:sprite}B). The co-regulation in expression of Map2k6 and Kcnj2 has been experimentally reported in \cite{melo2020hi}. Together, these indicate that the co-regulation may be a result of the same enhancer. In summary, we demonstrated the reliability of SNALS in predicting genetic higher-order interactions, as well as the potential of using hyperedge prediction to fully evaluate the effect of higher-order genetic interactions on gene expression.

\vspace{-2mm}
\section{Conclusion}

In this paper, we present SNALS framework to predict higher-order interactions in hypergraph. SNALS tackles the challenges caused by edge- and node-level ambiguities, which are unique to hyperedge prediction, but not in link prediction of ordinary graph. In doing so, SNALS utilizes bipartite graph neural network with structure features that collectively handle both ambiguities. Moreover, SNALS retrieved the spectrum information of the structure features,  which reflects the joint interaction between the hyperedge and its local environment. Such information could not be easily learned under the framework of current graph neural network. As a result, SNALS achieved nearly 30\% performance increase over most recent models. We further applied SNALS on predicting higher order genetic interaction. SNALS achieved consistent performance across different chromosomes and generated a highly plausible 4-way genetic interaction validated by existing literature. Such results further advocate the usefulness and applicability for SNALS in the task of hyperedge prediction.


\bibliography{neurips_2021}
\bibliographystyle{abbrv}

\newpage
\appendix

\section{Benchmark Datasets}
We evaluated our methods on eight datasets\footnote{Downloaded from https://www.cs.cornell.edu/~arb/data/}.

\begin{itemize}
    \item \textbf{DAWN}: Patient drug use recoreded in emergency room visits.
    \item \textbf{Email-Eu}: Emails with multiple email addresses.
    \item \textbf{NDC-classes}: Drugs with multiple classification labels.
    \item \textbf{NDC-substances}: Drugs consist of multiple substances.
    \item \textbf{Threads-ask}: Threads of users asking and answering questions on askubuntu.com
    \item \textbf{Threads-math}: Threads of users asking and answering questions on math.stackexchange.com.
    \item \textbf{Tags-ask}: Questions with multiple tags on askubuntu.com.
    \item \textbf{Tags-math}: Questions with multiple tags on math.stackexchange.com.
\end{itemize}

\begin{table}[ht]
  \label{sample-table}
  \begin{adjustbox}{max width=\linewidth,center}
  \begin{tabular}{lllllllll}
    \toprule
   \backslashbox{Statistic}{Data} & DAWN & email-Eu & NDC-class & NDC-substance & threads-ask & threads-math & tags-ask & tags-math \\
    \midrule
    No. Edge  & 138742 &  24399 & 1047 & 6264 & 115987 & 535323 & 145053 & 169259 \\
    No. Node & 2290 & 979& 1149& 3438& 90054 & 153806 & 3031& 1627 \\
    Edge degree & 3.987(2.207) & 3.488(2.849) & 6.115(4.839) & 7.964(5.910) & 2.309(0.635) & 2.610(0.933) & 3.427(0.992) & 3.497(0.945) \\
    Node degree & 241.554(1055.753) & 86.935(114.531) & 5.572(15.708) & 14.510(42.724) & 2.974(21.754) & 9.087(91.405) & 164.558(606.784) & 363.801(1040.086)\\

    \bottomrule
  \end{tabular}
  \end{adjustbox}
\caption{Dataset statistics, for edge and node degree, we report the mean value along with its standard derivation.}
\end{table}

Detailed statistics of the eight datasets are summarized in table 1. We also report the mean value of edge and node degree along with its standard derivation. We argue these datasets represent different scenarios in hypergraphs, including sparse (NDC-classes, threads-ask), medium (NDC-substance, threads-math), and dense (DAWN, email-Eu, tags-ask, tags-math) hypergraphs. We believe these datasets form a comprehensive benchmark set to will evaluate the performance and robustness of each model.

\section{Normalization on bipartite graph neural network}
The non-linear activation function in bipatite graph neural network captures the non-linear dependency of hyperedge with different edge-degrees, which introduce additional flexibility to the edge-embedding \(\bm{X}_E\) than the clique expansion based GNNs. Bipartite graph neural network is capable for representing hyperedge with different edge size. One important step in the bipartite graph neural network is to normalize node and edge embedding by their degree or size. Essentially, such normalizations balance the local topological characteristics and degree bias in embedding a single node or edge.  Noted, an over-normalization could eliminate contextual meaningful topological characteristics while none or less normalization causes a degree or size bias, i.e., the difference of embedding of nodes and edges is not in agreement with its topological characteristics but heavily influenced by its node degree or edge size. To test the impact of different levels of normalization on the model performance, we test the following four normalization scenarios:

\begin{align*}
&\text{Scenario 1:} \quad \bm{X}_E^{l+1}=\sigma(\bm{H}^T\bm{X}_V^l\bm{W}_E^l) \quad and \quad \bm{X}_V^{l+1}=\sigma(\bm{H}\bm{X}_E^l\bm{W}_V^l)\\
&\text{Scenario 2:} \quad \bm{X}_E^{l+1}=\sigma(\bm{H}^T\bm{X}_V^l\bm{D}_E^{-1}\bm{W}_E) \quad and \quad \bm{X}_V^{l+1}=\sigma(\bm{H}\bm{X}^l_E\bm{W}_V)\\
&\text{Scenario 3:}\quad \bm{X}_E^{l+1}=\sigma(\bm{H}^T\bm{X}_V^l\bm{W}_E) \quad and \quad \bm{X}_V^{l+1}=\sigma(\bm{D}_V^{-1/2}\bm{H}\bm{X}^l_E\bm{D}_V^{-1/2}\bm{W}_V)\\
&\text{Scenario 4:}\quad \bm{X}_E^{l+1}=\sigma(\bm{H}^T\bm{X}_V^l\bm{D}_E^{-1}\bm{W}_E) \quad and \quad \bm{X}_V^{l+1}=\sigma(\bm{D}_V^{-1/2}\bm{H}\bm{X}^l_E\bm{D}_V^{-1/2}\bm{W}_V)
\end{align*}

Specifically, scenario 1 corresponds to none normalization on both node and edge, which relies on \(\bm{W}_E\) and \(\bm{W}_V\) to compensate the degree impact. Scenario 2 and 3 that correspond to conducting the normalization on only edge-side or node-side, respectively. And scenario 4 normalizes both edge- and node-side. We compared the impact of the four normalization scenarios on SNALS on the benchmark datasets by fixing all other parameters. 

\begin{table}[h]
  \label{sample-table}
  \begin{adjustbox}{max width=\linewidth,center}
  \begin{tabular}{lcccccccc}
    \toprule
   \backslashbox{Method}{Data} & DAWN & email-Eu & NDC-class & NDC-substance & threads-ask & threads-math & tags-ask & tags-math \\
    \midrule
    Scenario 1 & 0.945(0.006) &  0.930(0.013) & 0.933(0.013) & 0.967(0.007) & 0.920(0.008) & 0.893(0.007) & 0.945(0.011) & 0.959(0.013) \\
    Scenario 2 & \textbf{0.956(0.011)} & \textbf{0.940(0.004)} & \textbf{0.957(0.010)} & \textbf{0.964(0.012)}& \textbf{0.906(0.013)} & \textbf{0.888(0.010)} & \textbf{0.941(0.014)}& \textbf{0.969(0.007)} \\
    Scenario 3 & 0.941(0.006) & 0.939(0.005) & 0.936(0.014) & 0.956(0.012) & 0.867(0.041) & 0.880(0.013) & 0.914(0.026) & 0.943(0.026) \\
    Scenario 4 & 0.950(0.006) & 0.939(0.005) & 0.933(0.019) & 0.960(0.019) & 0.857(0.049) & 0.874(0.017) & 0.928(0.017) &  0.967(0.008)\\

    \bottomrule
  \end{tabular}
  \end{adjustbox}
\caption{AUC results of different normalization scenarios for different hypergraph data.}
\end{table}

We report the AUC results of different normalization scenarios for different hypergraph data in table 2. Compared with none (scenario 1), node-side (scenario 3) and two-side normalization (scenario 4), edge-side normalization (scenario 2) consistently shows a better performance in all the eight benchmark datasets. Empirically, we argue that the one-side normalization would better balance the information loss and degree bias, such that it outperforms scenario 1 and 4. For the better performance of scenario 2 than scenario 3, we speculate a major reason is that we utilize the node-embedding rather than edge embedding to predict hyperedge. By omitting the normalization on node-side, the pipeline would take advantage of node embedding difference for a better prediction. Such that, in SNALS, we utilize the edge-size normalization scheme for the updating of node and edge embedding. We also noticed other works that introduce latent parameters to control the normalization \cite{dong2020hnhn}. This framework could certainly integrated in further improvement of SNALS.

\section{Set neural network for the standardization of structure features}
The structure feature \(\bm{X}_{S_{(q)}}\) characterizes the within hypergraph dependency of each node. By implementing structure feature within the bipartite graph neural network, \(f_{SN}\) collectively tackles both edge- and node-level ambiguities. One key component or requirement for \(f_{SN}\) is that \(\bm{X}_{S_{(q)}}\in \mathbb{R}^{|V_{(q)}|\times |S|}\) have different dimensions for different hyperedge \(S\). However, the bipartite graph neural network requires a fixed feature matrix \(\bm{X}_{V_{(q)}}\) as the input. To fill this gap, we utilize set neural network (setNN), precisely deepsets \cite{zaheer2017deep} to standardize \(\bm{X}_{S_{(q)}}\in\mathbb{R}^{|V_{(q)}|\times |S|}\) into \(\bm{X}_{V_{(q)}}\in\mathbb{R}^{|V_{(q)}|\times k}\), i.e., for each row of \(\bm{X}_{S_{{(q)}_{i:}}}\), we transfer it to a uniformed vector \(\bm{X}_{V_{{(q)}_{i:}}}\). Because of permutation invariant property of \(S\), any row-wise operation on \(\bm{X}_{S_{(q)}}\) should also be permutation invariant. SetNN fits this property perfectly as it regards \(\bm{X}_{S_{{(q)}_{i:}}}\) as a set rather than an ordered vector. Moreover, most setNN models like deepsets are very efficient to train and apply. One important parameter of setNN is the choice of pooling methods. Theoretically, any permutation invariant pooling methods (max-/mean-/sum-pooling) would maintain the permutation invariant property of setNN \cite{zaheer2017deep}. As for the case of hyperedge prediction, we recommend using sum-pooling which could reflect the edge--size information better than max or mean pooling. We also report their differences in table 3, as expected, sum-pooling (bold character) enjoys better and stable performance compared with max and mean pooling.

\begin{table}[h]
  \label{sample-table}
  \begin{adjustbox}{max width=\linewidth,center}
  \begin{tabular}{lcccccccc}
    \toprule
   \backslashbox{Method}{Data} & DAWN & email-Eu & NDC-class & NDC-substance & threads-ask & threads-math & tags-ask & tags-math \\
    \midrule
    max & 0.888(0.008) &  0.862(0.019) & 0.865(0.029) & 0.819(0.047) & 0.875(0.010) & 0.893(0.013) & 0.885(0.024) & 0.943(0.016) \\
    mean & 0.936(0.021) & 0.930(0.013) & 0.930(0.013) & 0.947(0.015) & 0.844(0.051) & 0.892(0.008) & 0.877(0.025) & 0.932(0.019) \\
     sum & \textbf{0.950(0.007)} & \textbf{0.939(0.007)} & \textbf{0.936(0.011)} & \textbf{0.956(0.013)}& \textbf{0.880(0.050)} & \textbf{0.899(0.007)} & \textbf{0.925(0.025)}& \textbf{0.955(0.009)} \\

    \bottomrule
  \end{tabular}
  \end{adjustbox}
\caption{AUC results of different pooling methods for deepsets.}
\end{table}

\section{The joint connection between hyperedge and its local environment.}
One key contribution of SNALS is the integration of spectrum information (\(f_{LS}\)) of the local topological characteristics in predicting hyperedges. As listed in the main text, our experiments demonstrated introducing \(f_{LS}\) consistently improved the model performance on all benchmark datasets, especially when GNN-based models did not performed well. Since the singular values represent the low rank property of the affinity matrix, \(f_{LS}\) characterizes the local topological characteristics of a node set when predicting if they form a hyperedge.

To further justify the necessity of \(f_{LS}\), we ask if the spectrum information could be directly learned by the existing neural network models, i.e., if we change the setting of our bipartite neural network model \(f_{SN}\) by a model including similar structure features,  would the enhanced \(f_{SN}\) alone achieves the same performance as SNALS(\(f_{SN}+f_{LS}\)). To this end, we replaced the deepsets module with set transformer \cite{lee2019set}, which is capable to learn higher order interactions within the set. Such property corresponds well with the joint interaction between hyperedge with its local environment. To evaluate, we also constructed four scenarios of different setNN with and without \(f_{LS}\): 1) deepsets, 2) deepsets \(f_{LS}\), 3) settransformer, 4) settransformer \(f_{LS}\). 

Here we report the F1 results of these scenarios in predicting hyperedge. For both setNN, the inclusion of \(f_{LS}\) maintains or strengthens the overall performance. Esepecially in the case that \(f_{SN}\) alone could not deliver a satisfactory performance (threads-ask, threads-math). We next focus on the comparison between settransformer and deepsets \(f_{LS}\). We regard the former one as learned higher order interactions and the later one as retrieved higher order interactions. In most of the cases, deepsets plus \(f_{LS}\) outperforms settransformer. This result suggest that to directly learn the higher order interaction by changing neural network structure is unlikely to match the performance of the integrating the spectrum information \(f_{LS}\). We then integrate set transformer with \(f_{LS}\) to test whether the combined effect could achieve better performance. Surprisingly, in many cases, these two information contradict with each other, which resulted a poor performance compared with settransformer or deepsets \(f_{LS}\) (DAWN, email-Eu, threads-ask, threads-math). Overall, this experiment revealed that higher order information is a nontrivial task for \(f_{SN}\). Even we include more advanced framework, like set transformer, the information learned may not necessarily reflect the true property of the hyperedge. This result further advocate the necessity of \(f_{LS}\) for the representation of hyperedge. Noted, overall deepsets plus \(f_{LS}\), i.e., the SNALS pipeline achieved best performance compared with all other settings.

\begin{table}[h]
  \label{sample-table}
  \begin{adjustbox}{max width=\linewidth,center}
  \begin{tabular}{lcccccccc}
    \toprule
   \backslashbox{Method}{Data} & DAWN & email-Eu & NDC-class & NDC-substance & threads-ask & threads-math & tags-ask & tags-math \\
   \midrule
    deepsets & 0.840(0.012) & 0.780(0.010) & 0.880(0.021) & 0.914(0.007) & 0.623(0.024) & 0.627(0.018) & 0.823(0.009) & 0.869(0.006) \\
    deepsets \(f_{LS}\) & 0.838(0.010) & 0.785(0.011) & 0.896(0.020) & 0.918(0.006) & 0.714(0.019) & 0.654(0.027) & 0.822(0.012) & 0.869(0.006)\\
    settransformer & 0.834(0.011) & 0.780(0.011) & 0.912(0.009) & 0.910(0.009) & 0.647(0.040) & 0.594(0.042) & 0.819(0.008) & 0.866(0.007) \\
    settransformer \(f_{LS}\) & 0.831(0.011) & 0.776(0.008) & 0.915(0.011) & 0.913(0.012) & 0.673(0.023) & 0.594(0.018) & 0.819(0.008) & 0.868(0.004)\\
    
    \bottomrule
  \end{tabular}
  \end{adjustbox}
\caption{F1 results of different methods in learning joint connections.}
\end{table}

\section{Other parameters for SNALS}
In practice, we only consider one-hop neighbors of the hyperedge. As shown in table 1, most hypergraph are dense graphs, hence the one-hop neighbor would balance the overall performance and computational cost. For deepsets and set transformer, we retrieved the code from \cite{lee2019set} and only modified the output dimension as 20, i.e., \(\bm{V}_{S_{(q)}}\in \mathbb{R}^{|V_{(q)}|\times|S|}\rightarrow \bm{X}_{V_{(q)}}\in \mathbb{R}^{|V_{(q)}|\times 20}\). The bipartite graph neural network has three identical layers, each with a node\(\rightarrow\)edge and edge\(\rightarrow\)node linear transformation followed by a ReLU activation. For a fair comparison with the strongest baseline method setSEAL, \(f_{SN}\) takes the same sortpooling approach in aggregating the information \cite{zhang2018link}. Currently, due to the varied size of hyperedge (i.e. \(|S|\geq2\)), we take the first two singular values as the input of \(f_{LS}\) to keep the uniformity of SNALS for hyperedges of different sizes. In our training procedure, we set our batch size as 50 and applied dropout. Throughout the experiments, we set at maximum 30 epoches for SNALS and used the default setting for all other methods. The detailed codes and toy examples are under preparation and will be released after paper acceptance.


\end{document}